# Scalable Distributed Job Processing with Dynamic Load Balancing

Srinivasrao Putti[1], Dr V P C Rao[2], Dr. A. Govardhan[3], Ambika Prasad Mohanty[4]

[1]Dept. of Computer Science, Y P R College of Engineering and Technology, AP, India
`yourpsr@gmail.com`
[2]Dept. of Computer Science, St. Peters Engineering College, AP, India
`pcrao.vemuri@gmail.com`
[3]Dept. of Computer Science, JNTU, Hyderabad, AP, India,
`govardhan_cse@yahoo.co.in`
[4]Infotech Enterprises, Hyderabad, AP, India,
`apmohanty@yahoo.com`

*ABSTRACT*

*We present here a cost effective framework for a robust scalable and distributed job processing system that adapts to the dynamic computing needs easily with efficient load balancing for heterogeneous systems. The design is such that each of the components are self contained and do not depend on each other. Yet, they are still interconnected through an enterprise message bus so as to ensure safe, secure and reliable communication based on transactional features to avoid duplication as well as data loss. The load balancing, fault-tolerance and failover recovery are built into the system through a mechanism of health check facility and a queue based load balancing. The system has a centralized repository with central monitors to keep track of the progress of various job executions as well as status of processors in real-time. The basic requirement of assigning a priority and processing as per priority is built into the framework. The most important aspect of the framework is that it avoids the need for job migration by computing the target processors based on the current load and the various cost factors. The framework will have the capability to scale horizontally as well as vertically to achieve the required performance, thus effectively minimizing the total cost of ownership.*

## 1. INTRODUCTION

The need for a distributed processing system arises from the fact that smaller and inexpensive heterogeneous computer systems should be utilized to achieve the required computation without a need for a large super computer. Such systems are usually independent with their own memory and storage resources, but connected to a network so that the systems communicate with each other for sharing the load. In such a computing environment, the systems usually remain idle until they are instructed to perform a computational task by a centralised monitor. Since the capabilities of such systems may vary, the central monitor usually keeps track of the load on each such system and assigns tasks to them. Over a period of time, the performance of each system may be identified and the information can be used for effective load balancing. Such distributed systems are extremely suitable for job processing. For load balancing, apart from the computational efficiency of each node, other factors like network latency, I/O overhead, job arrival rate, processing rate may be considered to distribute the jobs to various nodes so as to derive maximum efficiency and minimum wait time for jobs.

Various algorithms have been proposed for load balancing in distributed job processing systems. The algorithms can be classified into Static and Dynamic. While, the Static algorithm relies on a predetermined distribution policy, the Dynamic Load balancing algorithm makes its decisions based on the current state of the system. This framework uses the dynamic algorithm to analyse the current system load and various cost factors in arriving at the best target processor to handle the job processing.

## 2. RELATED WORK

Scheduling plays an important role in distributed systems in which it enhances overall system performance metrics such as process completion time and processor utilization [2]The basic idea behind distributed process scheduling is to enhance overall system performance metrics [4].Load sharing allows busy processors to load some of their work to less busy, or even idle, processors [5].Load balancing is a special case of load sharing, in which the scheduling algorithm is to keep the load even (or balanced) across all processors [6].Scheduling algorithms themselves can also be characterized as being either static or dynamic [2].

Static load balancing policies [20, 22, 23] use only the statistical information on the system (e.g., the average behaviour of the system)in making load-balancing decisions, and their principal advantage is lower overhead cost needed to execute them and their simplicity in implementation. Dynamic load balancing policies [8, 21, 22] attempt to dynamically balance the workload reflecting the current system state and are therefore thought to be able to further improve the system performance. Thus, it would be thought that, compared to static ones, dynamic load balancing policies are better able to respond to system changes and to avoid those states that result in poor performance. Load balancing policies can be classified as centralized or decentralized.

In centralized policies [17,19], it may be considered as a system with only one load balancing decision maker. The decentralized policies, on the other hand, delegates job distribution decisions to individual nodes . The decentralized load balancing is widely used to handle the imperfect system load information [19]. Sandeep Sharma, SarabjitSingh, and Meenakshi Sharma [2008] have studied various Static(Round Robin and Randomized Algorithms, Central Manager Algorithm, Threshold Algorithm) and Dynamic (Central Queue Algorithm, Local Queue Algorithm) Load Balancing Algorithms in distributed system and their results shows static load balancing algorithms are more stable . Vishwanathan developed incremental balancing and buffer estimation strategies to derive optimal load distribution for non critical load

In the traditional Sender / Receiver model[3] (we will call each such component as a node), a node acts as either a Sender or a Receiver. Each one has its own Job queue. Based on the prevailing load level crossing the threshold values, either Sender changes itself to a Receiver or Receiver changes itself to a Sender. In order that such a system works correctly, each such node needs to have the knowledge of all other nodes. This is the biggest disadvantage of such a system. This burdens the sender / receiver of having to discover other nodes by sending a broadcast request and expecting a response. When all these nodes do the same operation, there is considerable network overhead involved.

We proposed a Dispatcher / Processor model that is distinctly different from the Sender / Receiver model. The dispatcher has the responsibility of identifying a processor and dispatching a job. The processor will only execute the job.

we simulate the framework with a Java and JMS compliant ActiveMQ based monitor, dispatchers, processors and a centralized database. The framework will have the capability to scale horizontally as well as vertically to achieve the required performance, thus effectively minimizing the total cost of ownership.

## 3. APPROACH

In this article, we will discuss about an approach and feasible implementations of a priority and cost based job processing system with monitoring and load balancing capability. It is assumed that there are one or more processors available, but not necessarily online. By 'not necessarily online', we mean that a processor is capable of processing the job, but is currently not available and will be available in the near future. The system also has the reporting capability through its own persistence, possibly through a local or remote database. So, the status of a job is maintained in the persistence medium, a database. Since the data about the jobs is available centrally, load distribution can easily be supported. The reporting can be done from the data available in this database located centrally. With suitable monitoring and feedback capabilities, an intelligent Load balancing algorithm can be implemented.

Additionally, the framework allows the user to choose various algorithms through a set of configurable parameters, viz. based on Priority only, Cost only or Time based. With all these options, the scheduling remains dynamic in nature. When the user chooses Priority based scheduling, the scheduler identifies the best processor with minimal load that can handle the requested priority. If the user chooses Cost based scheduling, the scheduler, along with the current load of the system, takes into account various cost factors to arrive at the best processor that has minimal load and optimal cost effectiveness. The user can also choose a mixed mode where the scheduling is done with optimal load factor and cost.

The components (i.e. Dispatcher, Scheduler, Processor and Monitor) communicate over message queues, using a persistent message queue (part of Enterprise Message Bus). This solves the problem of sequencing of messages and avoids problems of messages being lost when the network fails of systems crash. The status of a job is maintained in the persistence layer, a database.

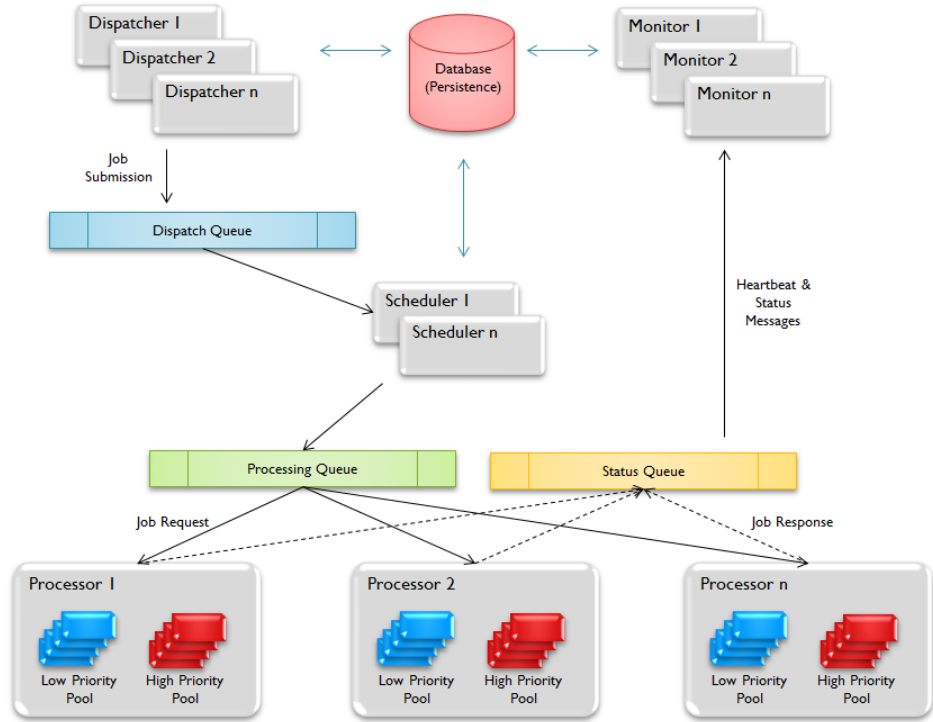

Figure 1: Architecture

## 4. DESIGN

The fundamental assumption in this design is the distributed nature of the nodes. The nodes may be present in any physical location, with any type of connectivity to a central message bus. The communication protocol used is standard TCP/IP. As will be clear later, the design has built-in load balancing option.

Most of the existing Sender/Receiver models need to have considerable knowledge of other receivers and have necessary logic to re-route a job. In the traditional Sender / Receiver model (we will call each such component as a node); a node acts as either a Sender or a Receiver. Each one has its own Job queue. Based on the prevailing load level crossing the threshold values, either Sender changes itself to a Receiver or Receiver changes itself to a Sender. In order that such a system works correctly, each such node needs to have the knowledge of all other nodes. This is the biggest disadvantage of such a system. This burdens the sender / receiver of having to discover other nodes by sending a broadcast request and expecting a response. When all these nodes do the same operation, there is considerable network overhead involved. Considerable amount of time is wasted at each node to query other nodes. This time could have been utilized for processing the job.

The proposed framework addresses these drawbacks and provides a better approach to managing the jobs. We propose a Dispatcher / Processor model that is distinctly different from the Sender / Receiver model. The dispatcher has the responsibility of identifying a processor and dispatching a job. The processor will only execute the job. Given below is brief description of various components of the system.

In order to simulate the model, two major components were developed using Java technology with ActiveMQ as messaging infrastructure. The first being the GridFramework and the second is the Grid Launcher.

## 4.1 COMPONENTS

- **Job Dispatcher** – This is the component that accepts the job requests (manual or otherwise), validates them and places the jobs in the Job Queue for scheduling. The dispatcher also records all the requests in the Database.

- **Job Scheduler** – This component receives a Job request from dispatcher, identifies the current load on the system and identifies the most suitable target processor that can process the new request. It then forwards the Job request to the target processor. Various options like Cost based, Priority based or Mixed mode can be specified with the job request so that the scheduler applies the appropriate algorithm to arrive at the best possible target processor for the job. However, it is possible to override this logic to enforce Processor affinity for a specific Job through suitable parameters for the job request.

- **Job Processor** – The processor is the component that picks up a job request from the queue, processes it. As shown in the diagram, the processor also reports the progress and status of job processing to the monitor. If a job is a long running job, progress information is sent at periodic intervals to the monitor. The Job processor also needs to report its health status back to the monitor.

- **Job Monitor** – This component is responsible for monitoring the status messages and updates the database. The component watches the progress messages and Heartbeat messages from various processors and saves the status in the database. This information also acts as feedback to the Job Dispatchers to take some intelligent decision at the time of dispatching the job to a target processor.

- **Dispatch Queue** – This is the message queue that stores the job requests dispatched until a processor picks them up for processing. Note that, for reliable job processing system, this Queue should have persistence capability, so that, in case of system failures, the requests lying in the queue are not lost.

- **Progress / Status Queue** – These are the message queues that store the job status sent by either dispatcher or processor. The monitor continuously monitors this queue for Job Status as well as Processor status messages. The information should include the current load, job status etc. This information is gathered by the Monitor and made available to the Job Dispatcher. The Job Dispatcher can then take intelligent decision based on this information to decide if a new job is to be dispatched to a target Job Processor or al alternate processor.

- **Database / Persistence** – This is the most critical component in the entire system. All the information about the Job, the Processors, the state of processing and the availability of processors are maintained at a central database. The proposed system also takes into account important design aspects that greatly enhance the Job processing. They are: Processor Affinity & Priority Thread Pool

- **Thread Pool** – We introduce here another important component in our design. The processor is designed to have a pool of threads. Each pool has a priority assigned to it.

Therefore, all the threads that are part of this group will inherit the priority assigned to the pool. Currently, the system supports two priority levels. They are: Low Priority and High Priority. However, the system has the flexibility to support more priority levels.

## 4.2 SOLUTION

Let us consider that all the processors are capable of handling any job and of any priority. In such a scenario, the system will have the following features:

  i.   Dispatcher can dispatch a job to the request queue, without bothering about the priority.
  ii.  The scheduler will have the capability to identify a processor based on the algorithm selected.
  iii. The processor is capable of handling jobs of any priority.
  iv.  The processor internally, maintains independent thread-pools for different priority jobs.
  v.   Based on the priority, the processor assigns the job to appropriate pool.
  vi.  The threads in a given pool have pre-defined priority, i.e. they are allocated CPU time based on the priority number assigned to them.

This solution appears simple and feasible. Let us discuss in detail about how such a system can be implemented.

## 4.3 ALGORITHMS

### 4.3.1 DISPATCH ALGORITHM

```
READ Job Definition From DATABASE
PREPARE Message
ASSIGN Job Request Options
DISPATCH Request
```

### 4.3.2 SCHEDULER ALGORITHM

```
READ Request
TargetNode = RequestNode
READ Alternate Targets From DATABASE
FOR EACH Alternate Target
   IF Target IS NOT AVAILABLE Continue To Next
   IF Least Load AND Target Load Is Minimum
      TargetNode = Target
      BREAK
   END IF
   IF Least Cost AND Target Cost Is Minimum
      TargetNode = Target
      BREAK
   END IF
END FOR
IF NO Target IS FOUND
   ABORT JOB
ELSE
   MARK TARGET FOR JOB AS TargetNode
   DISPATCH To TargetNode
ENDIF
```

### 4.3.3 PROCESSING ALGORITHM

```
WAIT For A Job Request
READ A Request
GET Job Priority

IF Priority = NORMAL
THEN
    ADD Job To Normal Priority Pool
ENDIF
IF Priority = HIGH
THEN
    ADD Job To High Priority Pool
ENDIF
UPDATE Job Status To IN-PROGRESS
```

### 4.3.4 MONITOR ALGORITHM

```
WAIT For A Status Message
READ A Message
GET Message Type
IF Message Type = HEARTBEAT
THEN
    UPDATE Processor Status
ENDIF
IF Message Type = JOB-STATUS
THEN
    UPDATE Job Status
ENDIF
```

### 4.4 ADVANTAGES OF THE PROPOSED SYSTEM

The advantages of such a Dispatcher / Processor model are as follows.

i. There is a clear distinction between Dispatcher (Sender), Scheduler, Processor (Receiver) and Monitor.
ii. Dispatcher just submits the Job request with appropriate parameters to the Scheduler.
iii. Processors are responsible for processing of the Jobs dispatched to them.
iv. When a job is submitted, the Scheduler analyses the status of all the processors and takes the intelligent decision about the best processor available.
v. As per Job definition, the dispatcher assigns priority to a Job.
vi. Processors have priority pools. A Job received is assigned to the respective pool, without interfering with the other Jobs being processed.
vii. Processors report status to a central monitor at a configurable interval and there is only one way communication.
viii. Scheduler need to query the central persistence (database) to check the status. This avoids nodes sending a request for status and other nodes responding with the status. This is a huge saving on the network usage.
ix. Any number of processors can be added and/or removed dynamically to the system without the need for configuration anywhere. Thus, the system has the ability to easily scale horizontally.

x. Each processor maintains it's internal Thread Pool based on the priority. The pool size is configurable. Thus, on a high end server, the same processor can be configured to handle more loads. This allows the system to easily scale vertically.
xi. Each Processor can be assigned an ID and thus, Processor affinity of a Job can be defined at the time of submission.
xii. Using a standard Message Queue with persistence helps the system retain the messages during a crash and subsequent recovery.
xiii. Processors utilize the time only for processing and need not have to be burdened with the decision of re-distributing the job when they are loaded. This situation will not arise because the dispatcher would have considered the load situation and distributed the job to the best processor which can immediately pick up the job for processing (assuming not all processors are 100% loaded). The job is dispatched only once. This minimizes wait of the jobs as well as network delays.
xiv. Theoretically, there is no limit on the number of threads in a priority pool. However, depending on the configuration of the hardware, processors may be configured to have appropriate priority pool sizes.

## 5. SIMULATION AND ANALYSIS

For simulating our design, we implemented a Java based job processing system with multiple processors, monitors and dispatchers. As part of this experiment, we defined a Job that compute 200,000 prime numbers. Thus, the Job processing time was allowed to take whatever time it takes to compute based on the priority assigned to the requested job. We started only one Processor and sent 100 Job requests. The wait time of the jobs were monitored

We performed experiments on the algorithm in three groups. The Group-1 experiment included one processor, one monitor, a dispatcher and a Scheduler. The Group-2 experiment included two processors, one monitor, a dispatcher and a Scheduler.

### 5.1 EXPERIMENT 1 (WITH NO TIME LIMIT)

We implemented a Java based job processing system with various options. As part of this experiment, we defined a Job that compute 200,000 prime numbers. Thus, the Job processing time was allowed to take whatever time it takes to compute.

i. The Job could take the priority as an attribute. The priority could be Low or High. Processors were assigned various cost factors for simulation.
ii. The scheduler was able to compute the most suitable target and dispatch the job to the appropriate queue (i.e. based on the options selected).
iii. The processor was designed to have two independent thread pools, one for Low priority and the other was for high priority.
iv. Each thread pool had the capacity to process 10 jobs concurrently, beyond which, jobs will wait in the queue.
v. When the Job processing was delegated to the appropriate thread, the thread priority was set to either low or high based on the Job's priority.
vi. The job was configured to compute 200,000 prime numbers.
vii. A total of 40 Jobs were dispatched, 20 with low priority and 20 with high priority.
viii. The wait time, processing time was measured for each job.
ix. Finally, the average values were plotted as a graph as shown in Figure 12.

Here is the data collected.

| Priority | Wait Time (ms) | Processing Time (ms) | Total Time (ms) |
|---|---|---|---|
| Low Priority | 231217 | 112141 | 343358 |
| High Priority | 133129 | 72955 | 206084 |

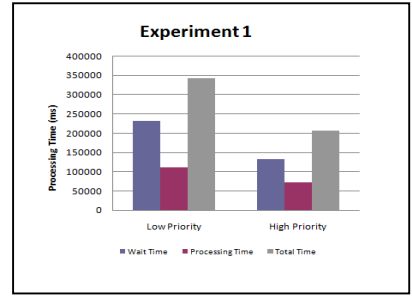

### 5.2 EXPERIMENT 2 (WITH TIME LIMIT)

We then implemented a Java based job processing system with more options. As part of this experiment, we defined a Job that compute prime numbers for a fixed period of approximately 40 seconds. Thus, the Job processing time was fixed and we monitored how many prime numbers were computed.

  i. The Job could take the priority as an attribute. The priority could be Low or High.
 ii. The scheduler was able to dispatch the job to the alternate queues (i.e. based on the option selected).

The processor was designed to have two independent thread pools, one for Low priority and the other was for high priority.

   i. Each thread pool had the capacity to process 10 jobs concurrently, beyond which, jobs will wait in the queue.
  ii. When the Job processing was delegated to the appropriate thread, the thread priority was set to either low or high based on the Job's priority.
 iii. The job was configured to compute for approximately 40 seconds.
  iv. A total of 40 Jobs were dispatched, 20 with low priority and 20 with high priority.
   v. The wait time, processing time and number of prime computations were measured for each job.
  vi. Finally, the average values were plotted as a graph as shown in Figure 13.

The data collected is shown in Table 4.

| Priority | Wait Time (ms) | Processing Time (ms) | Number of Primes |
|---|---|---|---|
| Low Priority | 104630 | 44014 | 349504 |
| High Priority | 116286 | 42012 | 701410 |

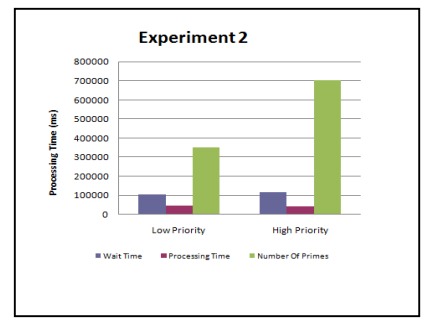

## 6. COMPARISON

We performed a comparison analysis of the data we collected after implementing a system using Sender initiated algorithm. The network overhead in the system was quite enormous and it increased the waiting time of the jobs as compared to that in the proposed algorithm. The results show that at-least 12% improvement in the total processing time in our proposed approach.

**Table 1**: Results

| Algorithm | Wait Time | Processing Time | Total Time |
|---|---|---|---|
| Sender | 284397 | 137933 | 422330 |
| Proposed Algorithm | 237549 | 134494 | 372043 |

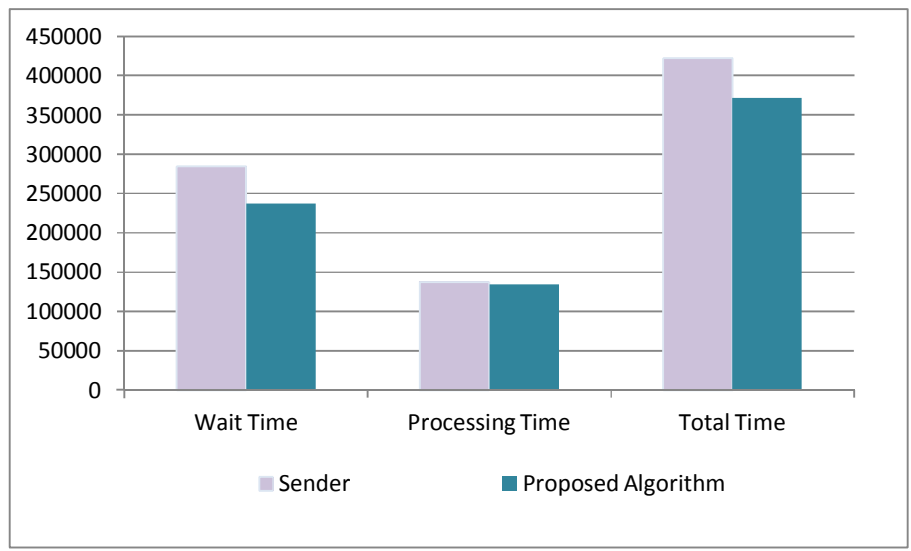

**Figure 2**. Comparison Results.

So far, in most of the systems implemented, the mechanism and protocol of communication between senders and receivers are not explained in detailed manner. This may lead to ambiguity in defining the overhead associated with the Sender initiated algorithms and/or Receiver initiated algorithms. The approach described here eliminates that ambiguity and also eliminates the overhead of processors participating in routing of jobs. This also keeps the architecture and implementation of such a system simple, dynamically scalable and flexible. This is an important aspect of requirement of a large array of networked Job processing systems.

## 7. CONCLUSIONS

The framework presented here can be easily implemented in a heterogeneous network of systems of varied capacity. The processors need not necessarily be of identical capability. Depending on the processing capacity of the systems, the processors can be configured to have, starting from 1 to any number of Threads with the required pool size and associated priority. The health and load of the processors in the network is available to any component in the network. The dispatchers running anywhere on the network can utilize this information for efficient routing.

As compared to the existing implementations, the framework is quite flexible and can be scaled up and scale out easily by changing few configuration parameters. As explained earlier, new jobs can be added easily by writing a job that implements the interface defined. Therefore, the expandability of the framework is quite high.

The future enhancements for improve reliability is to enhance failover-recovery mechanism for the processors. While, the experiment did not include the failover-recovery, the framework provides for maintaining the state of processing at various stages of processing. Therefore, adding a recovery mechanism will be quite easy by adding the feature of saving the state in a persistence medium and then recovering from where the processor failed after the next start-up.

Another enhancement would be to provide for Pause/Abort/Resume option for jobs. This feature would be of great benefit for long running Jobs in a network.

**AUTHORS**

**P Srinivasa Rao** did his B.Tech in 1998 from Osmania University, Hyderabad, India and M.Tech (Comp. Science) in 2003 from JNTU, Hyderabad, India. He has about 15 years of experience in teaching Computer Science at various engineering colleges. Currently, he is working as Professor and Principal at Y P R College of Engineering and Technology, AP, India. He specializes in Parallel and Distributed Systems. He is a life member of Indian Society for Technical Education (ISTE). His students have won several prizes at national level like IBM Great Mind Challenge. The author can be reached via e-mail: yourpsr@yahoo.com or yourpsr@gmail.com.

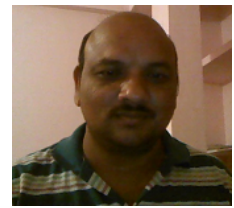

**Dr PC Rao** did his M.Tech and Ph.D. from IIT, Kharagur and he worked in IT industry for nearly 15 years and 10 years in teaching and as a professor. At present he is principal of Jyothishmathi Institute of Technological Sciences . He has published 20 papers in IEEE and INFOR, Philippine Statistician and International conferences. The author can be reached via e-mail: pcrao.vemuri@gmail.com.

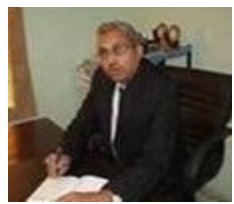

**Dr. A. Govardhan** did his B.E (CSE) in 1992, M.Tech in 1994 and Ph.D in 2003. He has published 125 technical papers in International and National journals and conferences, including IEEE, ACM and Springer. He has guided 6 PhDs, received several National and International Academic and service oriented awards. He has delivered several invited and keynote lectures. His areas is interest include Database, Data warehousing & Mining, Object Oriented Systems and Operating Systems, Currently, he is Professor in CSE & Director of Evaluation, JNTU Hyderabad AP, India. He has 18 years of experience in teaching and research. The author can be reached via e-mail: govardhan_cse@yahoo.co.in.

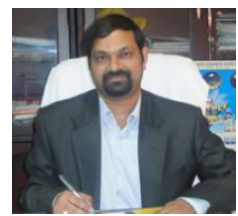

**Ambika Prasad Mohanty** did his B.Tech (Instrumentation) in 1988 from Madras Institute of Technology, Chennai, India, M.Tech (Computer Science) in 2003 from JNTU, Hyderabad, India and MBA (Technology Management) in 2011 from Anna University, TN, India. He had worked in the field of Instrumentation for four years and in the field of software design and development since 1993. The author specializes on software architectures, enterprise integration, and performance tuning, network and application security. He holds CVA and ITIL certification. The author can be reached via e-mail: apmohanty@yahoo.com.

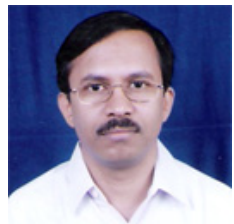